\documentstyle[amsfonts,epsfig,portland]{aipproc}
\begin{document}
\newcommand{\const}{\mbox{const} }
\newcommand{\Scri}{\mbox{$\cal J$}}
\newcommand{\E}[1]{{\cal E}_{\rm\bf #1\>}{}}
\newcommand{\N}[1]{{\cal N}_{\rm\bf #1\>}{}}
\newcommand{\Lien}{{\cal L}_{n}{}}
\newcommand{\DIII}{\nabla}
\newcommand{\tDIII}{\tilde{\nabla}}
\newcommand{\DeIII}{\Delta}
\newcommand{\RIII}{\,{}^{\scriptscriptstyle(3)\!\!\!\:}R}
\newcommand{\tRIII}{\,{}^{\scriptscriptstyle(3)\!\!\!\:}\tilde{R}}
\newcommand{\ROI}{\,{}^{\scriptscriptstyle(0,1)\!\!\!\:}\hat R}
\newcommand{\RII}{\,{}^{\scriptscriptstyle(1,1)\!\!\!\:}\hat R}
\newcommand{\epsIII}{\,{}^{\scriptscriptstyle(3)\!\!\!\:}\epsilon}
\newcommand{\hG}{\mbox{$\hat{G}$}}
\newcommand{\hR}{\mbox{$\hat{R}$}}
\newcommand{\hp}{\mbox{$\hat{\phi}$}}
\newcommand{\hT}{\mbox{$\hat{T}$}}
\newcommand{\Om}{\mbox{$\Omega$}}
\newcommand{\tC}{\mbox{$\tilde{C}$}}
\newcommand{\tG}{\mbox{$\tilde{G}$}}
\newcommand{\tT}{\mbox{$\tilde{T}$}}
\newcommand{\tg}{\mbox{$\tilde{g}$}}
\newcommand{\tn}{\mbox{$\tilde{\nabla}$}}
\newcommand{\tp}{\mbox{$\tilde{\phi}$}}
\newcommand{\tR}{\mbox{$\tilde{R}$}}
\newcommand{\thR}{\mbox{$\hat{\tilde R}$}}
\newcommand{\thT}{\mbox{$\hat{\tilde T}$}}
\newcommand{\thG}{\mbox{$\hat{\tilde G}$}}
\newcommand{\scri}{\mathcal{J}^+}
\newcommand{\Mt}{\widetilde{M}}
\title{Semiglobal Numerical Calculations of Asymptotically Minkowski
                 Spacetimes}
\author{Sascha Husa}
\address{Max-Planck-Institut f\"ur Gravitationsphysik\\ 
 Albert-Einstein-Institut \\
D-14476 Golm, Germany\\}
\maketitle
\begin{abstract}
 This talk reports on recent progress toward the {\em semiglobal}
 study of asymptotically flat spacetimes within numerical relativity.
 The development of a 3D solver for asymptotically Minkowski-like
 hyperboloidal initial data has rendered possible the application of
 Friedrich's conformal field equations to astrophysically interesting
 spacetimes.  As a first application, the whole future of a hyperboloidal
 set of weak initial data has been studied, including future null and
 timelike infinity. Using this example we sketch the numerical techniques
 employed and highlight some of the unique capabilities of
 the numerical code.
 We conclude with implications for future work. 
\end{abstract}

The modern treatment of gravitating isolated systems
allows to study asymptotic phenomena by local differential geometry.
The principal underlying idea, pioneered by Penrose \cite{Pe63ap}, is
to  work on an unphysical spacetime obtained from the
physical spacetime by a suitable conformal compactification.
Friedrich has extended this idea in a series of papers \cite{FriedrichOverview}
to the level of the field equations by rewriting Einstein's equations in a
regular way in terms of equations for geometric quantities on the
{\em unphysical spacetime} ${\cal M}$.
A metric $g_{ab}$ on ${\cal M}$ which is a solution
of the conformally rescaled equations gives rise to a physical metric
$\tilde g_{ab} = \Omega^{-2} g_{ab}$, where the conformal factor
$\Omega$ is also determined by the equations. The physical
spacetime $\tilde {\cal M}$ is then given by $\tilde {\cal M} =
\{ p \in {\cal M} \, \vert \, \Omega(p) > 0  \}$.
These ``conformal field equations'' render possible studies of the
global structure of spacetimes, e.g. reading off radiation at null
infinity, by solving regular equations.
It is natural to utilize everywhere spacelike slices $\Sigma_t$ in
${\cal M}$ which cross null infinity.
On $\tilde {\cal M}$ the corresponding slices $\tilde\Sigma_t$
are similar to the hyperboloid $t^2 - x^2 - y^2 - z^2 = k^2$ in Minkowski
spacetime, and are therefore usually referred to as hyperboloidal
slices. They are only Cauchy surfaces for the {\em future}
domain of dependence of initial slice of $\tilde {\cal M}$, we therefore
call our studies {\em semiglobal}.

The conformal field equations address a number of problems with the numerical
treatment of isolated systems in general relativity:
Radiation quantities can only be defined consistently at null infinity
($\scri$). Artificial outer boundaries cause 
ambiguities and stability problems. Resolving the different
length scales of radiating sources and the asymptotic falloff is numerically
difficult.  
Our {\em compactified} grid is allowed to extend beyond the physical part of
the unphysical spacetime, the boundary thus can not influence
the physics of a simulation. No artificial cutoff at some large distance
is required to keep the grid finite and there is no necessity to
treat very large length scales (dominating the asymptotic falloff)
along with variations on small scales.
Including $\scri$ in the computational domain enables straightforward
extraction of radiation quantities involving only well-defined
operations without ambiguities.
The {\em symmetric hyperbolicity} of the implemented formulation of the
conformal field equations guarantees a well-posed
initial value formulation. 

The rest of this article is organized in three parts: First
the current technology of the solution of the constraints will serve as
an example of how some technical problems which are particular to the
conformal field equations have been solved successfully.
After outlining the evolution algorithm
we discuss the evolution of weak data as an example of a situation for
which the usage of the conformal field equations is ideally suited: 
the main difficulties of the problem are directly addressed and solved
by using the conformal field equations.
These two topics sum up the work of H\"ubner on 3D numerical relativity with
the conformal field equations (see \cite{PeterIII},\cite{PeterIV} and
references cited therein).
Finally, we discuss future perspectives
for handling strong field situations, where a number of problems appear which
are not directly addressed by the conformal field equations, but
where we believe that their use will prove beneficial.

The constraints of the conformal field equations (see Eq. (14) of Ref.
\cite{Hu99ht}) are regular equations on the whole conformal spacetime
${\cal M}$.
However, they have not yet been cast into some standard type of PDE system,
such as a system of elliptic PDEs. One therefore resorts to a method where
one first obtains data for the
Einstein equations -- the first and
second fundamental forms $\tilde h_{ab}$ and ${\tilde k}_{ab}$ induced
on $\tilde\Sigma$ by $\tilde g_{ab}$ with corresponding
Ricci scalar and covariant derivative denoted by
$\tRIII$ and $\tDIII_a$.
After extending this {\em subset} of data from $\bar\Sigma$ to $\Sigma$
the data are then completed by using the conformal constraints.
Here we restrict ourselves to a subclass of hyperboloidal slices where
initially ${\tilde k}_{ab}$ is pure trace,
${\tilde k}_{ab} = \frac{1}{3} {\tilde h}_{ab} \tilde k$.
The momentum constraint
$
  \tDIII^b {\tilde k}_{ab} - \tDIII_a \tilde k = 0  
$
then implies $\tilde k = \const \ne 0$. We always set $\tilde k > 0$.
In order to reduce the Hamiltonian constraint
$$
\label{HamilConstr}
  \tRIII + {\tilde k}^2 = {\tilde k}_{ab}{\tilde k}^{ab}
$$
to {\em one} elliptic equation of second order,
we use the standard Lichnerowicz ansatz 
$$
  \label{tildeh}
  {\tilde h}_{ab} = \bar\Omega^{-2} \phi^4 h_{ab}.
$$
The free ``boundary defining'' function $\bar\Omega$ is chosen to
vanish on a 2-surface ${\cal S}$ -- the boundary of $\bar\Sigma$
and initial location
of $\scri$ -- with non-vanishing gradient on ${\cal S}$.
The topology of  ${\cal S}$ is chosen as spherical for asymptotically
Minkowski spacetimes. 
Let $h_{ab}$ be a metric on $\Sigma$, with the only restriction
that its  extrinsic 2-curvature induced by $h_{ab}$ on ${\cal S}$ is pure
trace, which is required as a smoothness condition \cite{AnCA92ot}. 
With this ansatz ${\tilde h}_{ab}$ is singular at ${\cal S}$,
indicating that ${\cal S}$ represents an infinity.
The Hamiltonian constraint then reduces to the Yamabe equation for the
conformal factor $\phi$:
$$
  4 \, \bar\Omega^2 \DIII^a \DIII_a\phi
  - 4 \, \bar\Omega (\!\DIII^a \bar\Omega)(\!\DIII_a \phi)
  - \left( \frac{1}{2} \RIII \, \bar\Omega^2 + 2 \bar\Omega \DeIII\bar\Omega
           - 3 (\!\DIII^a \bar\Omega) (\!\DIII_a \bar\Omega) 
    \right) \phi
  = \frac{1}{3} {\tilde k}^2 \phi^5.
$$
This is an ``elliptic'' equation with a principal part which vanishes at
the boundary ${\cal S}$ for a regular solution. This determines the
boundary values as
$
9 (\nabla^a \bar{\Omega}) (\nabla_a \bar{\Omega})
  = {\tilde k}^2 \phi^4
$.
Existence and uniqueness of a positive solution to the Yamabe 
equation and the corresponding existence and uniqueness of regular data for
the conformal field equations using the approach outlined above have been
proven by Andersson, Chru\'sciel and Friedrich~\cite{AnCA92ot}.

The {\em complete} set of data for the conformal field equations
is obtained from the conformal constraints via algebra and differentiation.
This however involves divisions by the conformal factor
$\Omega= \bar\Omega  \phi^{-2}$, which vanishes at ${\cal S}$.
In order to obtain a smooth error for this operation,
the numerically troublesome application of l'Hospital's rule to
$g = f/\Omega$  is replaced by solving an elliptic equation of the type
$
   \DIII^a \DIII_a ( \Omega^2 g - \Omega f ) = 0 
$
for $g$.
For the boundary values $\Omega^2 g - \Omega f = 0$, the unique solution is
$g=f/\Omega$. For technical details see H\"ubner \cite{PeterIII}. 
The Yamabe equation for $\phi$
and the linear elliptic equations arising from the
division by $\Omega$ are solved by pseudo-spectral collocation (PSC) methods
(see e.~g.~\cite{QuV97NA}).
Fast Fourier transformations converting between the spectral and grid
representations are performed using the FFTW library \cite{FrJ98fa}.
Nonlinearity in the Yamabe equation is dealt with by a multigrid
Newton method (for details and references see~\cite{PeterIII}, the
resulting linear equations are solved with the AMG
library~\cite{RuS87am}, which implements an algebraic multigrid
technique.
The PSC method restricts the choice of gridpoints to be consistent with
a simple choice of basis functions. Thus ${\cal S}$
is  required to be a coordinate isosurface and
the elliptic equations are solved in spherical coordinates. The polar
coordinate singularities are taken care of
by only using regular quantities in the computation, i.e. Cartesian tensor
components, and by not letting any collocation points coincide with
coordinate singularities.
In order to extend the initial data to the
Cartesian time evolution grid on the extended hyperboloidal slice
$\Sigma$, the spectral representations are used, which define data even
outside of the physical domain where the constraints have been solved.
The constraints will be violated in the unphysical region, but since this
region is causally disconnected from the physical interior by $\scri$,
the errors in the physical region converge to zero with
the discretization order. For numerical purposes the coefficient functions
of the spectral representation are modified in the unphysical region.

Time evolution of the conformal field equations (in particular here this
is the system Eq. (13) of Ref. \cite{Hu99ht}) is carried out by a 4th order
method of lines with standard 4th order Runge-Kutta time evolution. Spatial
derivatives are approximated by a symmetric fourth order stencil.
To ensure stability, dissipative terms of higher order were added
consistently with 4th order convergence, as discussed in section 6.7 of Ref.
\cite{GuKA95TD}.
%
%
In the unphysical region of ${\cal M}$ near the boundary of the grid, a
``transition layer'' is used to transform the conformal Einstein equations to
simple advection equations.
A trivial copy at the outermost gridpoint yields a simple and stable 
outer boundary condition.

The main result so far, obtained by H\"ubner, is the evolution
of weak data which evolve into a regular point $i^+$ representing
future timelike infinity.
This result illustrates a theorem by Friedrich \cite{Fr86ot},
who has shown that
for sufficiently weak initial data there exists a 
regular point $i^+$ of $\cal M$.
The complete future of (the physical part of) the initial slice
was reconstructed in a finite number of computational time steps,
and the point  $i^+$ has been resolved within a {\em single} grid
cell.  For these evolutions
very simple choices of the gauge source function have proven sufficient:
a zero shift vector was used, the lapse was set to $N = \sqrt{\det h}$, and
the scalar curvature of the unphysical spacetime was set to zero.
The initial conformal metric is chosen in Cartesian coordinates as
\begin{eqnarray*}
    h & = &
      \left(\begin{array}{ccc}
        1 + \frac{1}{3} \bar\Omega^2 \left( x^2 + 2 y^2 \right)
          & 0 & 0 \\
        0 & 1 & 0 \\
        0 & 0 & 1
      \end{array}\right)
  \end{eqnarray*}
and the boundary defining function as
$
 \bar\Omega = \frac{1}{2} \left( 1 - \left(x^2+y^2+z^2\right) \right).
$
The extraction of physics is largely based on the integration of
geodesics concurrently with the evolution. This is carried out
with the same 4th order Runge-Kutta scheme that is used in the method of
lines.
Null geodesics along $\scri$ are used to construct a Bondi system at $\scri$
and to compute the news function and Bondi mass.
To illustrate the results we show the behavior of geodesics in the numerically
generated spacetime. First, in Fig. 1 we show three timelike geodesics
originating with different initial velocities at the same point
$(x_0,y_0,z_0)=(\frac{1}{2\sqrt{3}},\frac{1}{2\sqrt{3}},\frac{1}{2\sqrt{3}})$
meeting a generator of $\scri$ at $i^+$. 
Fig. 2 shows the oscillations induced by gravitational waves
in the zero velocity geodesic starting out at $(x_0,y_0,z_0)$ 
(Fig. (6) of Ref. \cite{PeterIV} shows the same plot using coordinate time
instead of proper time). 
%
\begin{figure}[htbp] 
\centerline{\epsfig{file=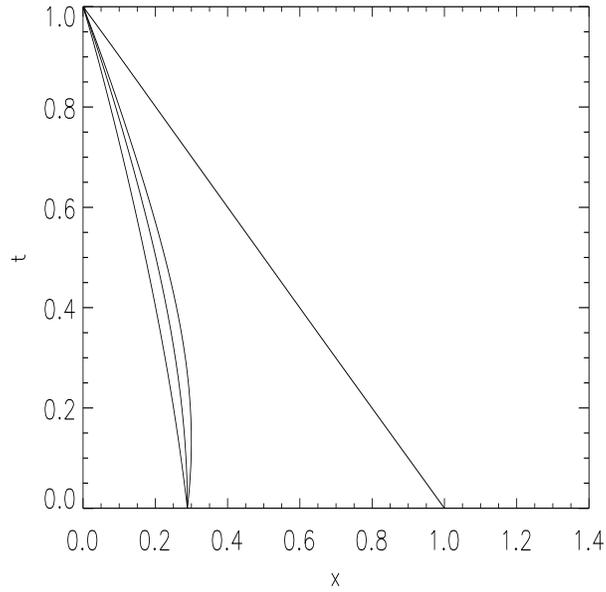,height=3.5in,width=3.5in}}
\vspace{10pt}
\caption{Three timelike geodesics 
 (starting at $x=\frac{1}{2\sqrt{3}}$) meet a generator of $\scri$
 (starting at $x=1$) at future timelike infinity $i^+$.}
\label{MeetingPoint}
\end{figure}
\begin{figure}[htbp] 
\centerline{\epsfig{file=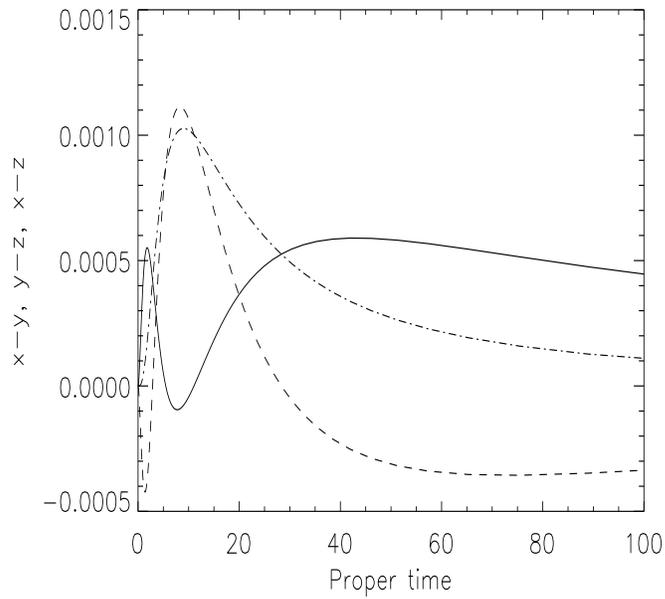,height=3.5in,width=3.5in}}
\vspace{10pt}
\caption{
The differences $x-y$, $x-z$,
and $y-z$ along a timelike geodesic show the spacetime
distortion due to gravitational waves.  
In flat space they would vanish for symmetry reasons.
The cutoff in time was chosen to better resolve the early time structure.}
\label{OscillateInProperTime}
\end{figure}

Future efforts will focus on developing the techniques to evolve
strong data without symmetries, e.g. to describe generic
black hole spacetimes. Here many problems are not yet solved, in particular
issues associated with choosing the gauge source functions,
the treatment of the appearance of singularities, and the limitations
of computer resources for 3D calculations. These well known
problems plaguing 3D numerical relativity will have to be addressed
and solved in the conformal approach in order to harvest its benefits
for studying the (semi)global structure of spacetimes.

A first step toward the study of black holes with the conformal
field equations is to obtain suitable initial data.
In the Cauchy approach topologically nontrivial data are
easy to produce by compactification methods (see e.g.
\cite{Hu98Bis}, \cite{Dain} and references cited therein)
where the topology of the computational grid is not influenced by the number
of asymptotic regions.
In the hyperboloidal case the adding of $\scri$'s as suggested e.g.
in \cite{PeterIII} {\em does} change the computational domain and leads to
technical problems, for which H\"ubner has suggested solutions \cite{PeterIII}.
An alternative is to use regular initial data containing one or
more apparent horizons. Such data could be produced by parameter studies
with the current code and would in some sense be more physical,
since they do not require the existence of ``eternal'' black holes, and in
principle should allow initial data which have apparent horizons and
singularities in the future, but not in the past.
An important question is whether such data are qualitatively different
from data with  ``topological'' black holes in the region outside of
the event horizon.
In order to handle the singularities inside of black holes with the code,
a strategy successfully employed by H\"ubner in the spherically
symmetric case was to handle floating point exceptions and then continue
past the formation of singularities as permitted by causality \cite{Hu96mf}.
However, the probably chaotic structure of the singularity and the difficulties
of evolving with a time step as large as permitted by causality may turn
out prohibitive in 3D.
Excision of the singularity or a modification of the equations inside the
black hole
(analogous to the way they are changed in the unphysical
region beyond $\scri$) are possible alternatives.

\section*{Acknowledgements}
The author thanks H. Friedrich, B. Schmidt and J. Winicour for helpful
discussions, and P. H\"ubner and M. Weaver for letting him use
their codes, explaining their results and providing general support in order
to take over this project.

\end{document}